\documentclass[reprint,aps,graphicx,prx,showpacs]{revtex4-1}
\usepackage{amsmath,amssymb,amsfonts}     

\usepackage{graphicx,bm,MnSymbol}
  \usepackage{paralist}
  \usepackage{epstopdf}
  \usepackage{graphics} 
 \usepackage[colorlinks=true]{hyperref}
 \hypersetup{urlcolor=blue, citecolor=red}
 \usepackage[latin1]{inputenc}
\usepackage[caption=false]{subfig}
\usepackage{soul}

\renewcommand{\theequation}{\arabic{section}.\arabic{equation}}

\newcommand{\e}{{\rm e}}

\newcommand{\calU}{\mathcal{U}}
\newcommand{\calG}{\mathcal{G}}
\newcommand{\calN}{\mathcal{N}}
\newcommand{\calS}{\mathcal{S}}

\newcommand{\p}{\widetilde{p}}

\newcommand{\calQ}{\mathcal{Q}}
\newcommand{\wcalQ}{\widetilde{\mathcal{Q}}}
\newcommand{\calP}{{\mathcal P}}
\newcommand{\calF}{{\mathcal F}}

\renewcommand{\P}{\mathbb{P}}
\newcommand{\R}{\mathbb{R}}

\newcommand{\E}{\mathbb{E}}

\renewcommand{\L}{{\mathbb L}}

\def\Z{\mathbb{Z}}
\def\P{\mathbb{P}}

\begin{document}

\title{Occupation time of a run-and-tumble particle with resetting}

\author{Paul C. Bressloff}
\address{Department of Mathematics, University of Utah, Salt Lake City, UT 84112 USA}

\begin{abstract}

We study the positive occupation time of a run-and-tumble particle (RTP) subject to stochastic resetting. Under the resetting protocol, the position of the particle is reset to the origin at a random sequence of times that is generated by a Poisson process with rate $r$. The velocity state is reset to $\pm v$ with fixed probabilities $\rho_1$ and $\rho_{-1}=1-\rho_1$, where $v$ is the speed. We exploit the fact that the moment generating functions with and without resetting are related by a renewal equation, and the latter generating function can be calculated by solving a corresponding Feynman-Kac equation. This allows us to numerically locate  in Laplace space the largest real pole of the moment generating function with resetting, and thus derive a large deviation principle (LDP) for the occupation time probability density using the Gartner-Ellis theorem. We explore how the LDP depends on the switching rate $\alpha$ of the velocity state, the resetting rate $r$ and the probability $\rho_1$. In particular, we show that the corresponding LDP for a Brownian particle with resetting is recovered in the fast switching limit $\alpha\rightarrow \infty$. On the other hand, the behavior in the slow switching limit depends on $\rho_1$ in the resetting protocol. 
 
 \end{abstract}

\maketitle

\section{Introduction}

Additive functionals provide important information concerning the spatio-temporal properties of the trajectory of a particle evolving according to a continuous stochastic process such as Brownian motion. If $X(t)$ denotes the position of the particle at time $t$, then an additive functional over a fixed time-interval $[0,T]$ is defined as a random variable $\calF_T$ such that
\begin{equation}
\label{eq1}
\calF_T=\int_0^T f(X(t))dt,
\end{equation}
where $f(x)$ is some prescribed function or distribution such that $\calF_T$ has positive support and $X(0)=x_0$ is fixed. Since $X(t)$ is a continuous stochastic process, it follows that each realization of a trajectory will typically yield a different value of $\calF_T$, which means that $\calF_T$ will be distributed according to some probability density function (pdf) $\calP(a,T|x_0,0)$. In the particular case of Brownian motion, the statistical properties of the associated Brownian functional can be analyzed using path integrals, and leads to the well-known Feynman-Kac formula \cite{Kac49}. That is,
let $\calG(k,t|x_0,0)$ be the moment generating function of $\calP(a,t|x_0,0)$:
\begin{equation}
\label{eq2}
\calG(k,t|x_0,0)=\int_0^{\infty}\e^{ka}\calP(a,t|x_0,0)da.
\end{equation}
Then $\calG$ satisfies the modified backward Fokker-Planck equation (FPE)
\begin{align}
\label{FK}
\frac{\partial \calG}{\partial t}=D\frac{\partial^2 \calG}{\partial x_0^2}+kf(x_0)\calG,
\end{align}
where $D$ is the diffusivity and $\calG(k,0|x_0,0)=1$. Brownian functionals are finding increasing applications in probability theory, finance, data analysis, and the theory of disordered systems \cite{Majumdar05}. Three additive functionals of particular interest are as follows \cite{Levy39,Ito} : (i) The occupation or residence time that the particle spends in $\R^+$ for which $f(x)=\Theta(x)$, where $\Theta(x)$ is the Heaviside function; (ii) The local time density for the amount of time that a particle spends at a given location $y$ for which $F(x)=\delta(x-y)$; (iii) The area integral obtained by setting $f(x)=x$.

Recently, a number of papers have explored the effects of stochastic resetting on the properties of additive functionals \cite{Meylahn15,Hollander19,Pal19}. Under a resetting protocol, the position of a particle is reset to some fixed point $x_r$ at a random sequence of times that is usually (but not necessarily) generated by a Poisson process with rate $r$. Following an initial study of Brownian motion under resetting \cite{Evans11a,Evans11b}, there has been an explosion of interest in the subject (see the recent review \cite{Evans20} and references therein). Much of the work has focused on random search processes and the observation that the mean first passage time (MFPT) to find a hidden target can be optimized as a function of the resetting rate. 
This phenomenon is particularly significant in cases where the MFPT without resetting is infinite, such as Brownian motion in an unbounded domain. Stochastic resetting renders the MFPT finite, with a unique minimum at an optimal resetting rate $0<r_{\rm opt} <\infty$. In a certain sense, resetting plays an analogous role to a confining potential.

The study of additive functionals with resetting is less developed. In Ref. \cite{Meylahn15} a renewal equation was derived that links the generating functions with and without resetting, analogous to the renewal equation linking the corresponding probability densities for the position $X(t)$ \cite{Evans20}. The derivation exploited the fact that when the particle resets, it loses all information regarding the trajectory prior to reset. Although the renewal formula applies to additive functionals of general Markov processes, concrete examples have to date been restricted to the area functional of an Ornstein-Uhlenbeck process \cite{Meylahn15}, the occupation time, area and absolute area functionals of Brownian motion \cite{Hollander19}, and the local time of a diffusing particle in a potential \cite{Pal19}. In each case, the authors investigated fluctuations of the relevant observable under resetting in the long-time limit, which can be characterized by the so-called rate function of large deviation theory \cite{Ellis85,Dembo98,Hollander00,Touchette09}. One of the interesting issues raised by these studies is to what extent observables that do not satisfy a large deviation principle (LDP) in the absence of resetting gain an LDP when resetting is included.  
This is analogous to the conversion of an infinite MFPT to a finite one due to the confining effects of resetting.

In this paper, we consider the effects of resetting on the occupation time of a run-and-tumble particle (RTP) that switches randomly between a left and right moving state of constant speed $v$. This type of motion arises in a wide range of applications in cell biology, including the unbiased growth and shrinkage of microtubules \cite{Dogterom93} or cytonemes \cite{Bressloff19}, the bidirectional motion of
molecular motors \cite{Newby10}, and the `run-and-tumble' motion of bacteria such as {\em E. coli} \cite{Berg04}. The run-and-tumble model has also attracted considerable recent attention within the non-equilibrium statistical physics community, both at the single particle level and at the interacting population level, where it provides a simple example of active matter \cite{Tailleur08,Cates15,Volpe16}. Studies at the single particle level include properties of the position density of a free RTP \cite{Martens12,Gradenigo19}, non-Boltzmann stationary
states for an RTP in a confining potential \cite{Dhar19,Sevilla19,Dor19}, first-passage time properties \cite{Angelani14,Angelani15,Malakar18,Demaerel18,Doussal19}, and RTPs under stochastic resetting \cite{Evans18}.

From a more general perspective, the motion of an RTP is governed by a symmetric, two-state version of a velocity jump process, which is itself an example of a piecewise deterministic Markov process (PDMP), also known as a stochastic hybrid system. Previously, we derived a general Feynman-Kac formula for additive functionals of a PDMP \cite{Bressloff17}, and used this to determine properties of the occupation time for a two-state velocity jump process, which included the RTP as a special case. Our results for the occupation time of an RTP were subsequently rediscovered in Ref. \cite{Singh19}. Here we combine our Feynman-Kac formulation of additive functionals for RTPs with the renewal approach of Ref. \cite{Meylahn15} in order to investigate the effects of resetting on the long-time behavior of the occupation time. 

We begin in section II by briefly reviewing the model of an RTP with resetting introduced in Ref. \cite{Evans18}. We highlight the fact that the resetting protocol also needs to specify a reset condition for the discrete velocity state, which is chosen to be consistent with renewal theory. In particular, we assume that the velocity state is reset to $\pm v$ with fixed probabilities $\rho_1$ and $\rho_{-1}=1-\rho_1$, where $v$ is the speed. We then calculate the nonequilibrium stationary probability density (NESS) and show that the bias in the reset protocol skews the density in the positive $x$-direction when $\rho_1>0.5$ and the negative direction when $\rho_1<0.5$. We also discuss the fast switching limit in which one recovers Brownian motion. In section III we define an additive functional for an RTP and briefly review the Gartner-Ellis theorem for LDPs, which relates the LDP rate-function to the largest real pole of the moment generating function in Laplace space. We then derive the renewal equation relating the moment generating functions with and without resetting along the lines of Ref. \cite{Meylahn15}, emphasizing the important role of the resetting protocol for the velocity state. We also write down the Feynman-Kac formula for the moment generating function without resetting, which was previously derived in the more general context of PDMPs \cite{Bressloff17}. A simplified version of the derivation is presented in appendix A. Finally, in section IV we apply the theory developed in previous sections to analyze the long-time behavior of the positive occupation time of an RTP with resetting. In particular, we explore how this depends on the resetting rate $r$, the switching rate $\alpha$ and the bias determined by $\rho_1$. We also compare our results to that of a Brownian particle with resetting, which is recovered in the fast switching limit.

\setcounter{equation}{0}
\section{Run-and-tumble particle with resetting}

Consider a particle that randomly switches between two constant velocity states labeled by $n=\pm$ with $v_+=v$ and $v_-=-v$ for some $v>0$. Furthermore, suppose that the particle reverses direction according to a Poisson process with rate $\alpha$. The position $X(t)$ of the particle at time $t$ then evolves according to the piecewise deterministic equation
\begin{equation}
\label{PDMP}
\frac{dX}{dt}=v\sigma(t),
 \end{equation}
where $\sigma(t)=\pm 1$ is a dichotomous noise process that switches sign at the rate $\alpha$. Following other authors, we will refer to a particle whose position evolves according to Eq. (\ref{PDMP}) as a run-and-tumble particle (RTP). Let $p_{\sigma}(x,t)$ be the probability density of the RTP at position $x\in \R$ at time $t>0$ and moving to the right ($\sigma=1)$ and to the left ($\sigma=-1$), respectively. The associated differential Chapman-Kolomogorov (CK) equation is then
\begin{subequations}
\label{DL}
\begin{align}
\frac{\partial p_{1}}{\partial t}&=-v \frac{\partial p_{1}}{\partial x}-\alpha p_{1}+\alpha p_{-1},\\
\frac{\partial p_{-1}}{\partial t}&=v \frac{\partial p_{-1}}{\partial x}-\alpha p_{-1}+\alpha p_{1}.
\end{align}
\end{subequations}
This is supplemented by the initial conditions $x(0)=x_0$ and $\sigma(0)=\sigma_0=\pm 1$ with probability $\rho_{\pm 1}$ such that $\rho_1+\rho_{-1}=1$.

The above two-state velocity jump process has a well known relationship to the telegrapher's equation. That is, differentiating Eqs. (\ref{DL}a,b) shows that
the marginal probability density $p(x,t)=p_0(x,t)+p_1(x,t)$ satisfies the telegrapher's equation \cite{Goldstein51,Balak88}
\begin{equation}
\left [\frac{\partial^2}{\partial t^2}+2\alpha \frac{\partial}{\partial t}-v^2\frac{\partial^2}{\partial x^2}\right ]p(x,t)=0. 
\end{equation}
(The individual densities $p_{0,1}$ satisfy the same equation.) One finds that the short-time behavior (for $t\ll \tau_c=1/2\alpha$) is characterized by wave-like propagation with $\langle x(t)\rangle^2\sim (vt)^2$, whereas the long-time behavior ($t\gg \tau_c$) is diffusive with $\langle x^2(t)\rangle \sim 2Dt,\, D=v^2/2\alpha $. For certain initial conditions one can solve the telegrapher's equation explicitly. In particular, if $p(x,0)=\delta(x)$ and $\partial_tp(x,0)=0$ then
\begin{align*}
&p(x,t)=\frac{e^{-\alpha t}}{2}[\delta(x-vt)+\delta(x+vt)]\\
&+\frac{\alpha\e^{-\alpha t}}{2v} \bigg [\Theta(x+vt)-\Theta(x-vt)\bigg ]\bigg[I_0(\alpha\sqrt{t^2-x^2/v^2})\\
&+\frac{t}{\sqrt{t^2-x^2/v^2}}I_0(\alpha\sqrt{t^2-x^2/v^2})\bigg ],
\end{align*}
where $I_n$ is the modified Bessel function of $n$-th order and $\Theta$ is the Heaviside function. The first two terms represent the ballistic propagation of the initial data along characteristics $x=\pm vt$, whereas the Bessel function terms asymptotically approach Gaussians in the large time limit. In particular $p(x,t)\rightarrow 0$ point wise when $t\rightarrow \infty$.

Now suppose that the position $X(t)$ is reset to its initial location $x_0$ at random times distributed according to an exponential distribution with rate $r\geq 0$ \cite{Evans18}. (For simplicity, we identify the reset state with the initial state.) We also assume that the discrete state $\sigma(t)$ is reset to its initial value $\sigma_0=\pm 1$ with probability $\rho_{\pm 1}$. The evolution of the system over the infinitesimal time $dt$ is then
\begin{subequations}
\label{X}
\begin{equation}
X(t+dt)=\left \{ \begin{array}{ccc}X(t)+v\sigma(t)dt & \mbox{with probability} & 1-rdt\\
x_0 & \mbox{with probability} & rdt,
\end{array} \right .
\end{equation}
and
\begin{equation}
\sigma(t+dt)=\left \{ \begin{array}{c}\sigma(t) \, \mbox{with probability} \, 1-rdt-\alpha dt\\
-\sigma(t) \, \mbox{with probability} \, \alpha dt   \\
\sigma_0=\pm 1 \, \mbox{with probability} \, r\rho_{\pm1}dt.
\end{array} \right .
\end{equation}
\end{subequations}
The resulting probability density with resetting, which we denote by $p_{r,n}$, evolves according to the modified CK equation \cite{Evans18} 
\begin{subequations}
\label{CKH2}
\begin{align}
\frac{\partial p_{r,1}}{\partial t}&=-v \frac{\partial p_{r,1}}{\partial x}-(\alpha +r) p_{r,1}+\alpha  p_{r,-1}
+r\delta(x-x_0)\rho_1,\\
\frac{\partial p_{r,-1}}{\partial t}&=v \frac{\partial p_{r,-1}}{\partial x}-(\alpha +r) p_{r,-1}+\alpha  p_{r,1}
+r\delta(x-x_0)\rho_{-1}.
\end{align}
\end{subequations}

In Ref. \cite{Evans18}, the NESS was determined in the symmetric case $\rho_{\pm}=1/2$ and $x_0=0$ by noting the the total density with resetting, $p_r= p_{r,1}+p_{-1}$, is related to the corresponding density without resetting, $p_0$, according to the renewal equation
\begin{align}
p_r(x,t)=\e^{-rt}p_0(x,t)+r\int_0^t\e^{-r\tau}p_0(x,\tau)d\tau.
\end{align}
The first term is the contribution from trajectories that do not reset, which occurs with probability $\e^{-rt}$, while the second term integrates the contributions from all trajectories that last reset at time $t-\tau$ (irrespective of their position). Taking the limit $t\rightarrow \infty$ implies that 
\begin{equation}
p_r^*(x)=\lim_{t\rightarrow \infty}p_r(x,t)=r\widetilde{p}_0(x,r),
\end{equation}
where $\widetilde{p}_0(x,r)$ is the Laplace transform of $p_0(x,t)$. The latter can be calculated by Laplace transforming Eqs. (\ref{DL}) and one finds that \cite{Evans18}
\begin{equation}
p_r^*(x)=\frac{\bar{\mu}_r}{2}\e^{-\bar{\mu}_r|x|},\quad \bar{\mu}_r=\sqrt{\frac{r(r+2\alpha)}{v^2}}.
\end{equation}

\begin{figure*}[t!]
\begin{center} 
\includegraphics[width=15cm]{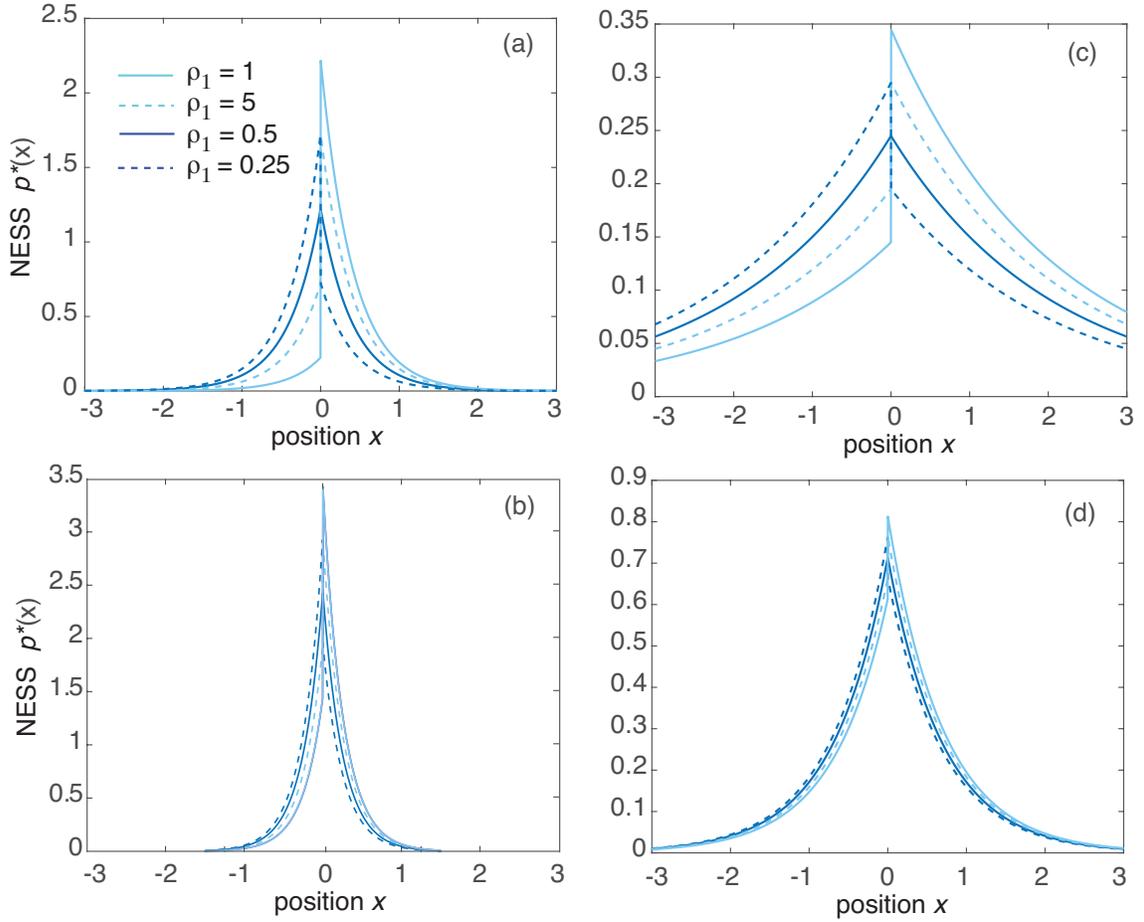} 
\caption{NESS $p^*(x)$ for an RTP that resets to the origin at a rate $r$; the velocity state is also reset to the initial value $\sigma v, \sigma \in \{-1,1\}$, with probability $\rho_{\sigma}$. Plots  are shown for various values of $\rho_1$ and (a) $\alpha =0.5,\ r=2$; (b) $\alpha=5,\ r=2$; (c) $\alpha=0.5,\ r=0.2$; (d) $\alpha=5,\ r=0.2$. The speed is taken to be $v=1$.}
\label{fig1}
\end{center}
\end{figure*}

However, in this paper, we are interested in how the occupation time of an RTP depends on the choice of resetting protocol. Therefore, we derive a more general expression for the NESS that applies for all choices of $\rho_{\pm 1}$. Again we take $x_0=0$. Rather than using a renewal equation, we directly solve Eqs. (\ref{CKH2}) using Laplace transforms. In Laplace space we have
\begin{subequations}
\label{CKL2}
\begin{align}
v \frac{\partial \p_{r,1}}{\partial x}&=-(\alpha +z+r) \p_{r,1}+\alpha  \p_{r,-1}+\frac{r+s}{s}\rho_1\delta(x),\\
-v \frac{\partial \p_{r,-1}}{\partial x}&=-(\alpha +z+r) \p_{r,-1}+\alpha  \p_{r,1}
+\frac{r+s}{s}\rho_{-1}\delta(x).
\end{align}
\end{subequations}
Differentiating either equation and rearranging one obtains a pair of decoupled second-order differential equations for all $x\neq0$:
\begin{equation}
\frac{d^2\p_{r,\pm 1}}{dx^2}-[(z+r+\alpha)^2-\alpha^2]\p_{r,\pm 1}=0,\quad x\neq 0.
\end{equation}
Requiring that the solution remains bounded as $x\rightarrow \pm \infty$ leads to the general solution
\begin{subequations}
\begin{align}
\p_{r,\pm 1}(x,z)&=A_{\pm} \e^{-\mu_r(z)x} \mbox{ for } x>0,\\ \p_{r,\pm1}(x,z)&=B_{\pm} \e^{\mu_r(z) x} \mbox{ for } x<0,
\end{align}
\end{subequations}
where
\begin{equation}
\mu_r(z)=\sqrt{\frac{(z+r)(z+r+2\alpha)}{v^2}}.
\end{equation}

We now need four algebraic conditions to determine the four coefficients $A_{\pm}$ and $B_{\pm}$. First,
substituting the general solution into the original first-order Eqs. (\ref{CKL2}) with $x\neq 0$ yields the conditions
\begin{subequations}
\begin{align}
(z+r+\alpha-\mu_r(z)v)A_+&=\alpha A_-,\\
 (z+r+\alpha+\mu_r(z)v)B_+&=\alpha B_-.
\end{align}
\end{subequations}
Second, integrating Eqs. (\ref{CKL2}) over the interval $x\in (-\epsilon,\epsilon)$ and taking the limit $\epsilon \rightarrow 0$ gives
\begin{equation}
\label{conda}
A_++A_--(B_++B_-)=\frac{r+s}{sv}(\rho_1-\rho_{-1}).
\end{equation}
Third integrating Eqs. (\ref{CKL2}) over $\R$ yields the conservation condition
\begin{align}
P=\int_{-\infty}^{\infty}(\p_1(x,z)+\p_{-1}(x,z))dx =\frac{1}{z},
\end{align}
which implies that
\begin{align}
\label{condb}
A_++A_-+B_++B_-=\frac{\mu_r(z)}{z}.
\end{align}
Finally, adding and subtracting conditions (\ref{conda}) and (\ref{condb})  we obtain the solution of the total probability density in Laplace space:
\begin{align}
&\p(x,z)=\p_{1}(x,z)+\p_{-1}(x,z)\nonumber \\
&=(A_++A_-)\e^{-\mu_r(z) x}\Theta(x)+(B_++B_-)\e^{\mu_r(z) x}\Theta(-x)\nonumber \\
&=\frac{\mu_r(z)}{2z}\e^{-\mu_r(z)|x|} +\mbox{sign}(x)\frac{r+z}{2zv}(\rho_1-\rho_{-1}) \e^{-\mu_r(z)|x|}.
\end{align}
Note that $\p(x,z)$ is discontinuous at $x=0$.
Using the result
\begin{equation}
p_r^*(x)=\lim_{z\rightarrow 0}z\p(x,z),
\end{equation}
and noting that $\mu_r(0)=\bar{\mu}_r$, the NESS is
\begin{align}
p^*(x)
&=\frac{\bar{\mu}_r}{2}\e^{-\bar{\mu}_r(|x|} +\mbox{sign}(x)\frac{r}{2v}(\rho_1-\rho_{-1}) \e^{-\bar{\mu}_r|x|}.
\end{align}
Clearly this reduces to the symmetric distribution when $\rho_1=\rho_{-1}=1/2$. On the other hand, the NESS is skewed towards positive (negative) values of $x$ when $\rho_1>\rho_{-1}$ ($\rho_1<\rho_{-1}$). This makes sense, since the particle is more likely to start off in one direction over the other, which adds a directional bias that is reinforced by resetting. On the other hand, the bias is reduced by increasing the switching rate $\alpha$ for fixed $r$. These various effects are illustrated in Fig. \ref{fig1}.

\setcounter{equation}{0}
\section{Generating functions and large deviations}

Let $\sigma_T=\{\sigma(t),\, 0\leq t \leq T\}$ denote a particular realization of the dichotomous noise process in the interval $[0,T]$. Let $X_{\sigma_T}(t)$ denote the corresponding solution of Eq. (\ref{X}a) and consider the functional 
\begin{equation}
\label{calT}
\calF_T=\int_{0}^T f(X_{\sigma_T}(t))dt,
\end{equation}
where $f$ is a real function or distribution. Here $\calF_T$ is a random variable with respect to different realizations of $\sigma_T$. Denote the corresponding probability density for $T^{-1}\calF_T$ (assuming it exists) by $\calP(a,T)$. Analogous to Brownian functionals, we will assume that in the large-$T$ limit, the probability density 
\[\calP(a,T)da=\P[a< T^{-1}\calF_T <a+da|X(0)=x_0]\]
has the form
\begin{equation}
\calP(a,T)=\e^{-TI_r(a)+o(T)},
\end{equation}
with $I_r(a)$ the so-called rate function.
This type of scaling is known as a large deviation principle (LDP) \cite{Ellis85,Dembo98,Hollander00,Touchette09}. It implies that the probability of observing fluctuations in $\calF_T$ at large times is exponentially small and centered about the global minimum of $I_r(a)$, assuming one exists. A typical method for determining the rate function of an LDP is to calculate the scaled cumulant function of $\calF_T$. The latter is defined as
\begin{equation}
\label{lam}
\lambda_r(k)=\lim_{T\rightarrow \infty}\frac{1}{T}\ln \E[\e^{k\calF_T}],
\end{equation}
where $k\in \R$ and $\E[\cdot]$ denotes the expectation with respect to different realizations $\sigma_T$, given that $X(0)=x_0$. If $\lambda_r(k)$ exists and is differentiable with respect to $k$, then one can use the Gartner-Ellis Theorem of large deviation theory, which ensures that $\calF_T$ satisfies an LDP with a rate function given by the Legendre-Fenchel transform of $\lambda(k)$ \cite{Ellis85,Dembo98,Hollander00,Touchette09}:
\begin{equation}
\label{GE}
I_r(a)=\sup_{k}\{ka-\lambda_r(k)\}.
\end{equation}

The quantity $\E[\e^{k\calF_T}]$ appearing in Eq. (\ref{lam}) is the scaled moment generating function of $\calP(a,T)$. That is,
\begin{equation}
\E[\calF_T^n]=\left . \frac{\partial^n}{\partial k^n}\E[\e^{k\calF_T}]\right |_{k=0}.
\end{equation}
In Ref. \cite{Meylahn15}, renewal theory is used to derive an integral equation that expresses the moment generating function with resetting in terms of the corresponding moment generating function without resetting. Although the authors focus on SDEs, they highlight the fact that their analysis also carries over to other Markov processes. Here we applytheir derivation to an RTP with resetting. It is useful to include the details of the analysis in order to highlight the fact that one also has to specify a reset rule for the discrete variable $\sigma(t)$. Let
\begin{equation}
\label{Qn}
\calG_{r}(x_0,t,k)=\E[\e^{k\int_{0}^t f(X_{\sigma_t}(s))ds}].
\end{equation}
be the generating function for the RTP with resetting, which evolves according to Eqs. (\ref{X}a,b). Assume that over the time interval $[0,T]$ there are ${\mathcal N}$ resettings with intervals $\tau_1,\ldots \tau_{\mathcal N}$ such that $T=\sum_{l=1}^{\calN +1}\tau_l$,
where $\tau_{\calN+1}$ is the time since the last resetting. The integral defining $\calF_T$ can then be partitioned into a sum of integrals:
\begin{equation}
\calF_T= \sum_{l=1}^{\calN +1}\int_{T_{l-1}}^{T_{l-1}+\tau_l}f(X(s))ds,
\end{equation}
where $\tau_0=0$, $T_{l-1}=\sum_{j=1}^l \tau_{j-1}$, and $X(s)$ evolves according to Eq. (\ref{PDMP}) in each integral domain. In order to determine $\calU_{n}$, we have to sum over all possible reset events (number of events $\calN$ and their reset times). Since the probability density of having a reset at time $\tau$ is $r\e^{-r\tau}$ and the probability of no reset until time $\tau$ is $\e^{-r\tau}$, the moment generating function decomposes as \cite{Meylahn15}
\begin{align}
\label{ren}
&\calG_{r}(x_0,T,k)=\sum_{\calN=0}^{\infty}\int_0^Td\tau_1 r\e^{-r\tau_1}\calG_{0}(x_0,\tau_1,k)\\
&\times \int_0^Td\tau_2 r\e^{-r\tau_2}\calG_{0}(x_0,\tau_2,k)\nonumber \\
&\times \cdots  \int_0^Td\tau_{\calN+1} \e^{-r\tau_{\calN+1}}\calG_{0}(x_0,\tau_{\calN+1},k)\delta(T-\sum_{l=1}^{\calN +1}\tau_l),\nonumber
\end{align}
where $\calQ_{0}$ is the corresponding generating function without resetting.

The above renewal equation exploits the fact that each reset returns the system to its initial state $(x_0,\sigma_0)$ with $\sigma_0$ generated from the distribution $\rho_{\sigma_0}$. A standard method for solving such an equation is to use Laplace transforms. Let
\begin{equation}
\widetilde{\calG}_{r}(x_0,z,k)=\int_0^{\infty}\e^{-zT}\calG_{r}(x_0,T,k)dT.
\end{equation}
Assuming that we can reverse the summation over $l$ and integration with respect to $T$, we can Laplace transform each term in Eq. (\ref{ren}). For example, setting 
\[q(\tau)=\e^{-r\tau}\calQ_{0}(x_0,\tau,k),\]
we have
\begin{align*}
&\int_0^{\infty}\e^{-zT}  \int_0^Td\tau_1q(\tau_1)\int_0^Td\tau_2 q(\tau_2)\delta(T-\tau_1-\tau_2)\\
&=\int_0^{\infty}\e^{-zT}  \int_0^Td\tau_1q(\tau_1) q(T-\tau_1)=\widetilde{q}(z)^2
\end{align*}
from the convolution theorem, where $\widetilde{q}(z)=\widetilde{\calQ}_{0}(x_0,z+r,k)$. Hence,
\begin{align*}
&\widetilde{\calG}_{r}(x_0,z,k)\\
&=\widetilde{\calG}_{0}(x_0,z+r,k)\sum_{\calN=0}^{\infty}
r^{\calN}\widetilde{\calG}_{0}(x_0,z+r,k)^{\calN}.
\end{align*}
Assuming that $r\widetilde{\calG}_{0}(x_0,z,k)<1$, the geometric series can be summed to yield the result \cite{Meylahn15}
\begin{equation}
\label{ren2}
\widetilde{\calG}_{r}(x_0,z,k)=\frac{\widetilde{\calG}_{0}(x_0,z+r,k)}{1-r\widetilde{\calG}_{0}(x_0,z+r,k)}.
\end{equation}

From the definition of the generating function,  Eq. (\ref{lam}) can be rewritten as
\begin{equation}
\lambda_r(k)=\lim_{T\rightarrow\infty} \frac{1}{T}\ln\calG_{r}(x_0,z,k).
\end{equation}
This then implies that
\begin{equation}
\calG_{r}(x_0,T,k)\sim \e^{\lambda_r(k)T}
\end{equation}
as $T\rightarrow \infty$, so that
\begin{equation}
\widetilde{\calG}_{r}(x_0,z,k)\sim \frac{1}{z-\lambda_r(k)}.
\end{equation}
Hence, as for SDEs with resetting \cite{Meylahn15}, one can determine $\lambda_r(k)$ by identifying the largest simple and real pole of the right-hand side of Eq. (\ref{ren2}). The latter will correspond to a zero of $1-r\widetilde{\calG}_{r}$ when $\widetilde{\calG}_{r}$ is finite. Finally, if $\lambda_r(k)$ is differentiable, then we can obtain the rate function $I_r(k)$ for a PDMP with resetting by taking the Legendre-Fenchel transform of $\lambda_r(k)$.

In the case of SDEs, it is well-known that the generating function without resetting satisfies a Feynman-Kacequation \cite{Ellis85,Dembo98,Hollander00,Touchette09}. An analogous result holds for velocity jump processes without resetting. Introduce the conditional generating function 
\begin{equation}
\label{Qn0}
\calQ_{\sigma_0}(x_0,t,k)=\E[\e^{k\int_{0}^t f(X_{\sigma_t}(s))ds}|1_{\sigma(0)=\sigma_0}],
\end{equation}
with $X_{\sigma_t}(s)$ satisfying Eq. (\ref{PDMP}).  It follows that
\begin{equation}
\label{G0}
\calG_0(x_0,t,k)=\rho_1\calQ_1(x_0,t,k)+\rho_{-1}\calQ_{-1}(x_0,t,k)
\end{equation}
In appendix A, we use a modified version of the path-integral construction developed in Ref. \cite{Bressloff17} to show that $\calQ_{\sigma_0}$ evolves according to the Feynman-Kac equation
\begin{subequations}
  \label{swCK}
 \begin{eqnarray} 
  \frac{\partial \calQ_{1}}{\partial t}   &=&v \frac{\partial \calQ_{1}}{\partial x_0}+kf(x_0)\calQ_{1}-\alpha \calQ_{1}+\alpha \calQ_{-1},\\
 \frac{\partial \calQ_{-1}}{\partial t}   &=&-v \frac{\partial \calQ_{-1}}{\partial x_0}+kf(x_0)\calQ_{-1}-\alpha \calQ_{-1}+\alpha \calQ_{1}.\nonumber \\
  \end{eqnarray}
  \end{subequations}
  Laplace transforming this equation with respect to $\tau$ gives
  \begin{subequations}
 \label{swLCK}
  \begin{eqnarray} 
 -1 &=&v\frac{\partial \wcalQ_1}{\partial x_0}+kf(x_0)\wcalQ_1-(z+\alpha) \wcalQ_{1}+\alpha \wcalQ_{-1} ,\\
  -1 &=& -v\frac{\partial \wcalQ_{-1}}{\partial x_0 }+kf(x_0)\wcalQ_{-1}+\alpha \wcalQ_1-(z+\alpha) \wcalQ_{-1}.\nonumber \\
  \end{eqnarray}
  \end{subequations}
In the following we drop the subscript $0$ on $x_0$ in order to simplify the notation.

  \setcounter{equation}{0}
\section{Positive occupation time}

Suppose that $x\in \R$ and consider the occupation time defined by Eq. (\ref{calT}) with $f(x)=\Theta(x)$, where $\Theta(x)$ is the Heaviside function:
\begin{equation}
\calF_T=\int_{0}^T \Theta(X_{\sigma_T}(t))dt
\end{equation}
We first calculate the Laplace transformed generators $\wcalQ_{\pm 1}$ along the lines of Ref. \cite{Bressloff17}, and then use Eq. (\ref{ren2}) to determine the generator with resetting, $\calG_r$.
 For the given choice of $f(x)$, we have to solve Eqs. (\ref{swLCK}) separately in the two regions $x>0$ and $x<0$, and then impose continuity of the solutions at the interface $x=0$. In order to determine the far-field boundary conditions for $x\rightarrow \pm \infty$, we note that
if the system starts at $x=\pm \infty$ then it will never cross the origin a finite time $\tau$ in the future, that is, 
\[\calP(\calF_T,t|\infty,0)=\delta(t-\calF_T),\quad \calP(\calF_T,t|-\infty,0)=\delta(\calF_T).\]
 Substituting this into the definition of $\widetilde{Q}_n$ shows that
\begin{equation}
\label{far}
\wcalQ_n( \infty;z,k)=\frac{1}{z-k},\quad \wcalQ_n( -\infty;z,k)=\frac{1}{z}.\end{equation}
Therefore, setting 
\begin{align*}
\widetilde{Q}_n(x;z,k)&=u_n^+(x; z,k)+\frac{1}{z-k},\quad x>0,\\  \widetilde{Q}_n(x;z,k)&=u_n^-(x; z,k)+\frac{1}{z},\quad x<0,
\end{align*}
we have
\begin{subequations}
 \label{obackQ2}
  \begin{eqnarray} 
0 &=&v\frac{\partial u_1^+}{\partial x}-(z-k+\alpha)u_1^++\alpha u_{-1}^+, \\
  0 &=&-v\frac{\partial u_{-1}^+}{\partial x }+\alpha u_1^+-(z-k+\alpha) u_{-1}^+,
  \end{eqnarray}
  and
  \begin{eqnarray} 
0 &=&v\frac{\partial u_1^-}{\partial x}-(z+\alpha)u_1^-+\alpha u_{-1}^- ,\\
  0 &=&-v\frac{\partial u_{-1}^-}{\partial x }+\alpha u_1^--(z+\alpha) u_{-1}^-.
  \end{eqnarray}
  \end{subequations}
with corresponding boundary conditions $u^{\pm}_n(\pm \infty;z,k)=0$. Eqs. (\ref{obackQ2}) can be rewritten in the matrix form
\begin{equation}
\frac{\partial}{\partial x}\left (\begin{array}{c}u_1^+\\ u_{-1}^+\end{array} \right )+{\bf M}(z-k)\left (\begin{array}{c}u_1^+\\ u_{-1}^+\end{array} \right )=0,\quad x\in (0,\infty),
\end{equation}
and
\begin{equation}
\frac{\partial}{\partial x}\left (\begin{array}{c}u_1^-\\ u_{-1}^-\end{array} \right )+{\bf M}(z)\left (\begin{array}{c}u_1^-\\ u_{-1}^-\end{array} \right )=0,\quad x\in (-\infty,0),
\end{equation}
with
\begin{equation}
{\bf M}(z)=\left (\begin{array}{cc} \displaystyle -\frac{z+\alpha}{v} & \displaystyle \frac{\alpha}{v} \\ \\  \displaystyle -\frac{\alpha}{v} &\displaystyle \frac{z+\alpha}{v} \end{array} \right ).
\end{equation}

The matrix ${\bf M}(z)$ has eigenvalues
\begin{align}
\label{lamp}
\lambda_{\pm} (z)=\pm \frac{\sqrt{z^2+2\alpha z}}{v}.
\end{align}
The corresponding eigenvectors are 
\begin{equation}
\label{eig}
{\bf w}^{\pm}(z)=\left (\begin{array}{c} \displaystyle \frac{z+\alpha}{v}-\lambda_{\pm}(z) \\ \\ \displaystyle\frac{\alpha}{v}\end{array} \right ).
\end{equation}
In order that the solutions $u_n^{\pm}$ vanish in the limits $x\rightarrow \pm \infty$, they have to take the form
\begin{subequations}
\begin{align}
u_n^+(x;z,k)&=Aw_n^+(z-k) \e^{-\lambda_+(z-k)x},\quad x\in (0,\infty),\\
u_n^-(x;z,k)&=B w_n^-(z) \e^{-\lambda_-(z)x},\quad x\in (-\infty,0).
\end{align}
\end{subequations}
We thus have two unknown coefficients $A,B$, which are determined by imposing continuity of the solutions $\wcalQ_n^{\pm}$, $n=\pm$, at $x=0$. This yields the two conditions
\begin{subequations}
\label{fourc}
\begin{align}
Aw_1^+(z-k)+\frac{1}{z-k}&=Bw_1^-(z)+\frac{1}{z},\\
Aw_{-1}^+(z-k)+\frac{1}{z-k}&=Bw_{-1}^-(z)+\frac{1}{z}
\end{align}
\end{subequations}
Adding and subtracting these equations gives
\begin{align*}
AD_+(z-k)&=BD_-(z)\\
AS_+(z-k)&=BS_-(z)+\frac{2}{z}-\frac{2}{z-k},
\end{align*}
where
\[S_{\pm}(z)= w_1^{\pm}(z)+ w_{-1}^{\pm}(z),\quad D_{\pm}(z)=w_1^{\pm}(z)- w_{-1}^{\pm}(z).\]
Hence
\begin{subequations}
\begin{align}
A&=\left [S_+(z-k)-\frac{S_-(z)D_+(z-k)}{D_-(z)}\right ]^{-1}\left [\frac{2}{z}-\frac{2}{z-k}\right ],\\
B&=\left [\frac{S_+(z-k)D_-(z)}{D_+(z-k)}-S_-(z)\right ]^{-1}\left [\frac{2}{z}-\frac{2}{z-k}\right ]
\end{align}
\end{subequations}  

In the following we will assume that the initial (reset) position is $x_0=0$. It then follows from Eq. (\ref{G0}) that the Laplace transformed generating function without resetting is given by
\begin{align}
\label{GG0}
&\widetilde{\calG}_{0}(0,z,k)=\frac{1}{z}+\left [\frac{2}{z}-\frac{2}{z-k}\right ]
\\
&\times \frac{\rho_1\left [(z+\alpha)/v-\lambda_-(z)\right ]D_+(z-k)+\rho_{-1}\alpha D_-(z)/v}{S_+(z-k)D_-(z)-S_-(z)D_+(z-k)}\nonumber .
\end{align}
Substituting Eq. (\ref{GG0}) into Eq. (\ref{ren2}) then yields the corresponding generating function with resetting, $\widetilde{\calG}_{r}(0,z,k)$.

\begin{figure}[t!]
\begin{center} 
\includegraphics[width=8.5cm]{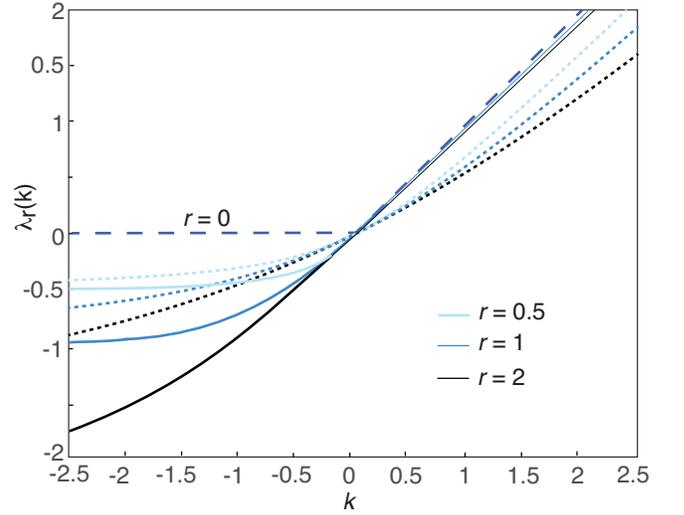} 
\caption{Largest real pole of the Laplace transformed generating function for the occupation time of an RTP that resets to the state $x_0=0$ and $\sigma_0=1$. Plot of $\lambda_r(k)$ as a function of $k$ for various resetting rates $r$. The solid curves are for the RTP and the dotted curves are for a corresponding Brownian particle with resetting. The dashed line indicates the dominant pole without resetting, $\lambda_0(k)$. Other parameter values are $v=1$ and $\alpha=0.5$.}
\label{fig2}
\end{center}
\end{figure}

\subsection{Principal pole $\lambda_r(k)$}

The poles of $\widetilde{\calG}_{r}(0,z,k)$ in the complex $z$-plane can be determined numerically and the largest real pole yields $\lambda_r(k)$ for a given $k$. Let us begin by considering the case that the particle always starts in the right-moving state, $\rho_1=1$. In Fig. \ref{fig2} we plot $\lambda_r(k)$ as a function of $k$ for various resetting rates $r$. In the absence of resetting we find that $\lambda_0(k)=k$ for $k>0$ and $\lambda_0(k)=0$ for $k\leq 0$ (dashed line in Fig. \ref{fig2}). In this case $\lambda_0(k)$ is not differentiable at $k=0$ and is not strictly convex, indicating that there does not exist an LDP. On the other hand, if $r>0$ then $\lambda_r(k)$ is a continuously differentiable and strictly convex function of $k$, consistent with the existence of an LDP. In addition, we find that the curves have the horizontal asymptotes $\lambda_r(k)\rightarrow -r$ as $k\rightarrow -\infty$, whereas $\lambda_r(k)<\lambda_0(k)$ for $k>0$. 

Also shown in Fig. \ref{fig2} are the corresponding plots for a Brownian particle with resetting, which was previously analyzed in \cite{Hollander19}. The latter authors used the well-known result that the Laplace transform of the generator in the absence of resetting takes the form \cite{Majumdar05}
\begin{align}
\label{BG}
\widetilde{\calG}_{0}(0,z,k)&= \frac{1}{\sqrt{z(z-k)}}.
\end{align}
This can be inverted to obtain an explicit expression for the so-called ``arcsine'' law for the probability density of the occupation time for pure Brownian motion starting at the origin \cite{Levy39} :
\begin{equation}
\calP(\calF_T=a,T)= \frac{1}{\pi\sqrt{a(T-a)}},\quad 0<a < T.
\end{equation}
As noted in Ref. \cite{Hollander19}, the non-exponential form of the arcsine law and the fact that $\calP(a,T)$ does not concentrate as $T\rightarrow \infty$ indicate that an LDP does not exists when $r=0$.
Substituting Eq. (\ref{BG}) into (\ref{ren2}) then gives
\begin{align}
\label{Br}
\widetilde{\calG}_{r}(0,z,k)&= \frac{1}{\sqrt{(z+r)(z+r-k)}-r}.
\end{align}
It follows that the poles of $\calG_r$ are determined in terms of solutions to the equation
\[(z+r)(z+r-k)=r^2,\]
which implies that the leading real pole is
\begin{equation}
\label{lala}
\lambda_k(r)=\frac{1}{2}\left [k-2r+\sqrt{k^2+4r^2}\right ].
\end{equation}
This formula determines the dotted curves in Fig. \ref{fig2}. Two major differences from the RTP curves are (i) they approach the horizontal asymptotes $-r$ much more slowly as $k\rightarrow -\infty$; (ii) they deviate more significantly from $\lambda_r=k$ when $k>0$. The behavior in the large-$|k|$ regime can be further identified by Taylor expanding the expression for $\lambda_r(k)$:
\begin{equation}
\lambda_r(k)\sim \frac{1}{2}\left [k-2r +|k|(1+2r^2/k^2)+O(r^3)\right ],
\end{equation}
which shows that $\lambda_r(k)\sim -r$ for $k \ll -1$ and $\lambda_r(k) \sim k-r$ for $k\gg r>0$.

\begin{figure}[t!]
\begin{center} 
\includegraphics[width=8.5cm]{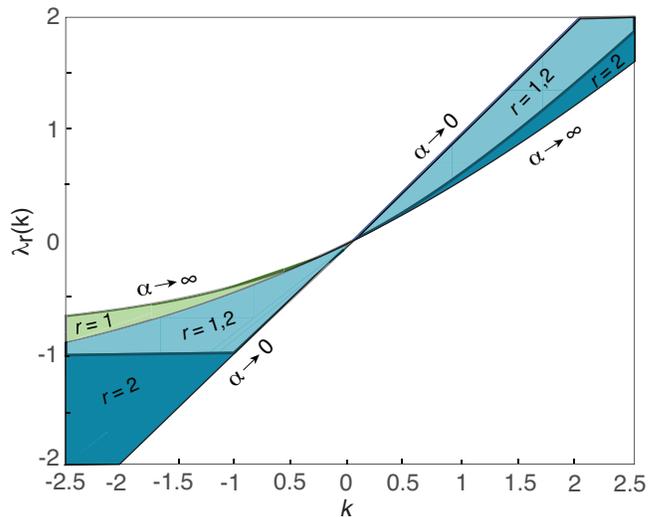} 
\caption{Dependence of $\lambda_r(k)$ on the switching rate $\alpha$ for $r=1,2$ and $\rho_1=1$ (particle always resets to the right-moving state). The shaded regions show the range of values of $\lambda_r(k)$ for a given $r$ as $\alpha$ is varied from zero to infinity. Other parameters are as in Fig. \ref{fig2}. }
\label{fig3}
\end{center}
\end{figure}

\begin{figure}[t!]
\begin{center} 
\includegraphics[width=8.5cm]{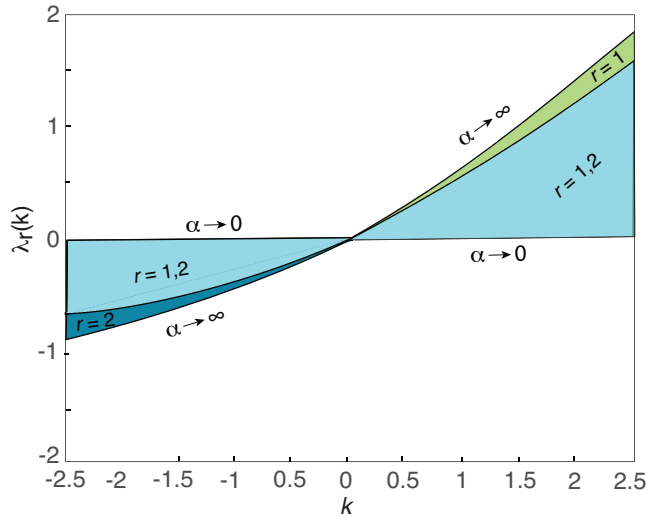} 
\caption{Same as Fig. \ref{fig3} except that $\rho_1=0$ (particle always resets to the left-moving state).}
\label{fig4}
\end{center}
\end{figure}

The differences between the RTP and Brownian particle vanish in the fast switching limit $\alpha \rightarrow \infty$, which is a consequence of the relationship between the CK equation of the RTP and the telegrapher's equation, see section II.
In particular, taking $k,z\ll \alpha$, we obtain the asymptotic behaviors $S_{\pm}(z)\rightarrow 2\alpha/v $ and
$D_{\pm}(z)\rightarrow \pm \lambda(z)$ with $\lambda(z)= \sqrt{2\alpha z}/v$. The leading order approximation of the coefficient $B$ is then
\begin{align}
B&\sim -\frac{v}{\alpha}\frac{\sqrt{z-k}}{(\sqrt{z-k}+\sqrt{z})}\left [\frac{1}{z}-\frac{1}{z-k}\right ]\\
&=\frac{v}{\alpha}\frac{\sqrt{z}-\sqrt{z-k}}{z\sqrt{z-k}},\end{align}
and the asymptotic solution for $\widetilde{\calG}_{0}(0,z,k)$ reduces to Eq. (\ref{BG}). This asymptotic result holds for all choices of the probability $\rho_1$ in the case of finite $r$. On the other hand, the behavior in the slow switching limit $\alpha\rightarrow 0$ is strongly dependent on $\rho_1$. For example, if $\rho_1=1$ as in Fig. \ref{fig2}, then the particle always starts out in the positive $x$ direction and rarely reverses its speed. This means that $\Theta(X(t))=1$ for almost all times $t$ and in the limit $\alpha\rightarrow 0$ we have 
$\calP(a,T)\rightarrow \delta(a-1)$ and $\widetilde{\calG}_{r}(0,z,k)\rightarrow 1/(z-k)$.
In Fig. \ref{fig3} we plot the range of values of $\lambda_r(k)$ for $r=1,2$ as $\alpha$ varies in the interval $(0,\infty)$. The $\alpha \rightarrow \infty$ boundaries coincide with the dotted curves of Fig. \ref{fig2}, whereas the $\alpha\rightarrow 0$ boundary is given by the straight line $\lambda_r(k)=k$. In Fig. \ref{fig4} we show the corresponding diagram in the case $\rho_1=0$. Now the particle always starts in the left-ward moving state so that $\Theta(X(t))=0$ for almost all times $t$, 
$\calP(a,T)\rightarrow \delta(a)$ and $\widetilde{\calG}_{r}(0,z,k)\rightarrow 1/z$. The zero $\alpha$ boundary is now the horizontal line $\lambda_r(k)=0$. (If $0<\rho_1<1$ then the $\alpha =0$ boundary is $\lambda_r(k)=0$ for $k<0$ and $\lambda_r(k)=k$ for $k>0$.)

\begin{figure}[b!]
\begin{center} 
\includegraphics[width=8.5cm]{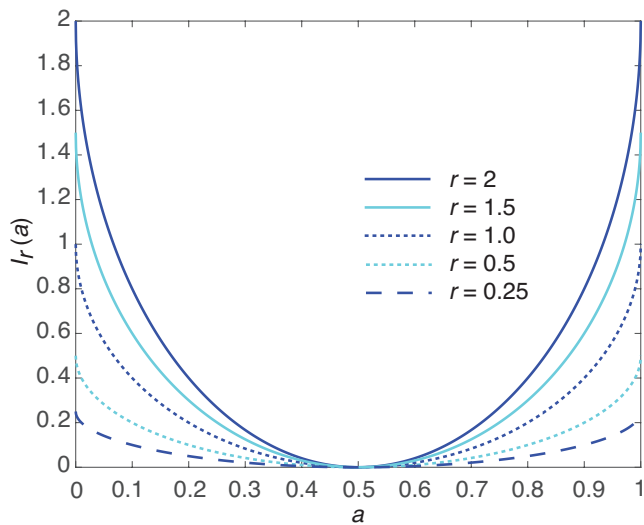} 
\caption{Plot of rate function $I_r(a)$ for the occupation time density in the case of a Brownian particle with resetting to the origin.}
\label{fig5}
\end{center}
\end{figure}

\begin{figure*}[t!]
\begin{center} 
\includegraphics[width=15cm]{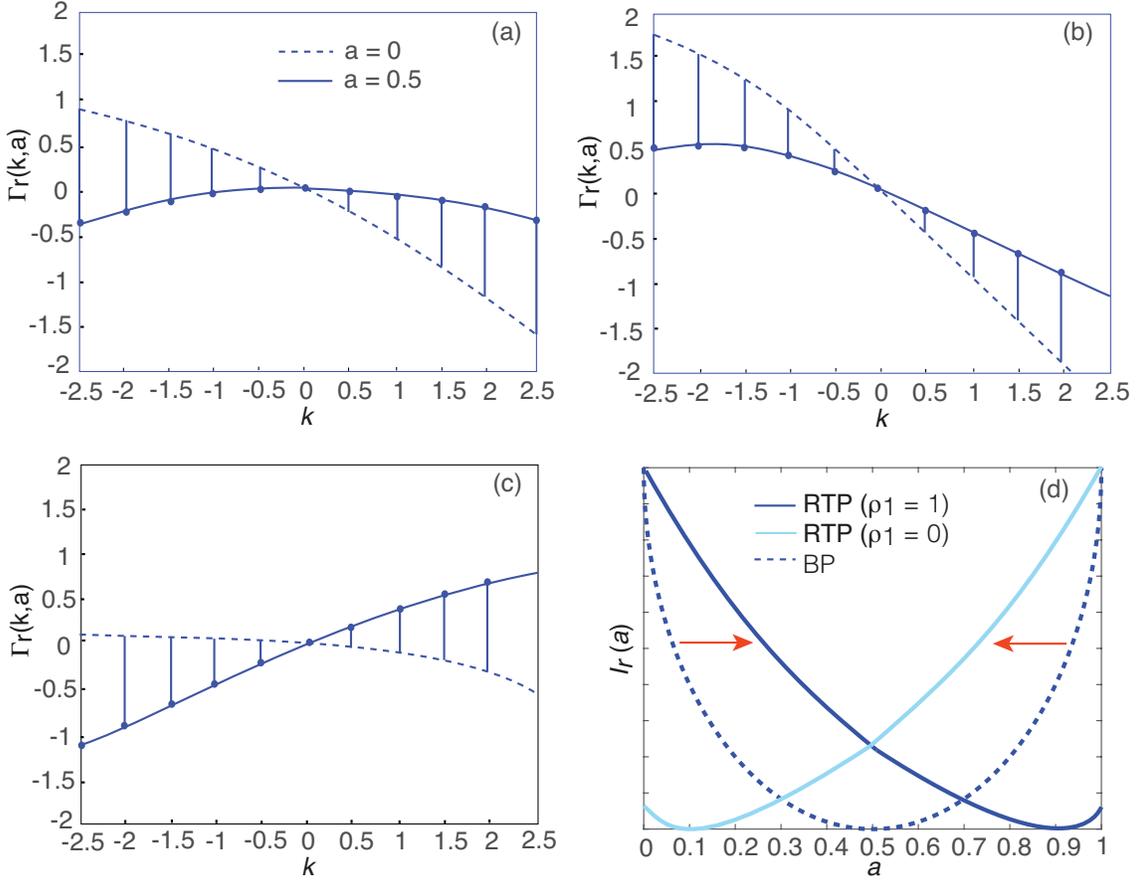} 
\caption{Graphical construction of the rate-function for the occupation time of an RTP with resetting. (a-c) Plots of the function $\Gamma_a(k)=ka-\lambda_r(k)$ as a function of $k$ for $a=0$ (dashed curve) and $a=0.5$ (solid curve) for (a) a Brownian particle, (b) an RTP with $\rho_1=1$, and (c) an RTP with $\rho_1=0$. The peak of the solid curves determines $I_r(a)$ for the given $a$. (d) Effect of $\rho_1$ on the rate function $I_r(a)$.  The shifts in the rate function curves increase with $r$ and decrease with $\alpha$. Other parameters are $r=2$, $\alpha=0.5$, and $v=1$.}
\label{fig6}
\end{center}
\end{figure*}

\subsection{Rate function $I_r(a)$}

Given a strictly convex, differentiable principal pole $\lambda_r(k)$ one can apply the Gartner-Ellis theorem to determine the rate function $I_r(a)$ of the LDP. First, consider the case of pure Brownian motion \cite{Hollander19}. Eq. (\ref{GE}) reduces to a simple Legendre transformation in which $a=\partial \lambda_r(k)/\partial k$.  From Eq. (\ref{lala}) we have
\begin{equation*}
a=\frac{1}{2}\left [1+\frac{k}{\sqrt{k^2+4r^2}}\right ],
\end{equation*}
which can be rearranged to give
\begin{equation}
k=k(a)=\frac{r(2a-1)}{\sqrt{(1-a)a}},\quad a\in (0,1).
\end{equation}
Hence,
\begin{equation}
I_r(a)=r(1-2\sqrt{a(1-a)}),\quad a\in (0,1).
\end{equation}
Note that $I(1/2)=0$ and $I(a)$ is strictly positive for $a\in [0,1/2)\bigcap(1/2,1]$. Since the Brownian particle dynamics is symmetric about the origin, the corresponding rate function is also symmetric with a minimum at $a=0.5$. That is in the long-time limit, the particle is expected to spend equal amount of times in the positive and negative domains so that the most likely value of the occupation time is $a=1/2$. The restriction of $a$ to the domain $[0,1]$ reflects the fact that $0\leq \calF_T/T\leq 1$.

Calculating the rate function in the case of an RTP has to be carried out numerically. However, the qualitative differences between the rate-functions of an RTP and a Brownian particle can be discerned using the graphical construction shown in Fig. \ref{fig6}. For a fixed value of $a$ and $r$, we vertically displace the curves $-\lambda_r(k)$ by $ka$. This generates the curve $\Gamma_r(k,a)=ka-\lambda_r(k)$ whose supremum with respect to $k$ determines $I_r(a)$ for the specific choice of $a$. Given the fact that the $\lambda_r(k)$ curves for the RTP are tilted in the clockwise (anticlockwise) direction around the origin relative to the corresponding curves for the Brownian particle when $\rho_1=1$ ($\rho_1=0$), the shift in the peak of $\Gamma_r(k,a)$ as a function of $a$ can be deduced. In particular there exists a crossover point $a=a_c$. In the case $\rho_1=1$, we find that $I^{\rm rtp}_r(a)> I_r^{\rm bp}(a)$ for $0<a<a_c$ and $I^{\rm rtp}_r(a)<I_r^{\rm bp}(a)$ for $a_c<a< 1$, where $a_c$ depends on $\alpha$ and $r$.  In particular, the rate function is no longer symmetric about $a=0.5$ and its minimum is shifted towards $a=1$. This is consistent with the observation that the NESS is also shifted to the right when $\rho_1>0.5$, see section II. In the slow switching limit $\alpha \rightarrow 0$, the density $\calP(T^{-1}\calF_T=a,T)\rightarrow \delta(a-1)$. Similarly, when $\rho_1=0$, the minimum of the rate function is shifted toward $a=0$. This is illustrated schematically in Fig. \ref{fig6}(d).

\section{Discussion}

In this paper we have used a mixture of renewal theory, large deviation theory and a Feynamn-Kac formula to investigate the long-time behavior of the occupation time of an RTP with stochastic resetting. We focused on how the behavior compared with a Brownian particle with resetting, which is obtained in the fast switching limit, and the dependence on the resetting protocol for the discrete velocity state. In future work we hope to extend our analysis to other additive functionals of RTPs with resetting. It would also be of interest to consider other examples of PDMPs, given that both the renewal equation (\ref{ren2}) and a Feynman-Kac formula (see Eq. (\ref{swCK2})) apply to this more general class of stochastic process. One simple extension would be to consider a directed velocity jump process with resetting, as recently studied in Ref. \cite{Bressloff20}. Now there is a directional bias when the reset protocol is unbiased.

\renewcommand{\theequation}{A.\arabic{equation}}

\setcounter{equation}{0}

\section*{Appendix A: Feynman-Kac operator for a PDMP without resetting}

In this appendix we simplify the path-integral construction of Ref. \cite{Bressloff17} in order to derive the Feynman-Kac operator of Eq. (\ref{swCK}). For the sake of generality, consider a system whose states are described by a pair
$(x,\sigma) \in \R \times \Gamma$, where $x$ is a continuous variable and $\sigma$ a discrete stochastic variable taking values in the finite set $\Gamma$ with $|\Gamma|=M$ When the internal state is $n$, the system evolves according to the ordinary differential equation (ODE)
\begin{equation}
\label{eq:deterministic}
\dot{x}=F_n(x),
\end{equation}
where $F_n: \R \to \R$ is a continuous function. For fixed $x$, the discrete stochastic variable evolves according to a homogeneous, continuous-time Markov chain with generator ${\bf A}(x)$. The generator is related to the transition matrix ${\bf W}$ of the discrete Markov process according to
\[A_{nm}=W_{nm}-\delta_{n,m}\sum_lW_{ln},\]
with $W_{mm}=0$ for all $m$.
We make the further assumption that the chain is irreducible for all $x\in \Sigma$, that is, for fixed $x$ there is a non-zero probability of transitioning, possibly in more than one step, from any state to any other state of the Markov chain. This implies the existence of a unique invariant probability distribution on $\Gamma$ for fixed $x\in \Sigma$, denoted by the vector ${\bf p}^*(x)$ with ${\bf p}^*=(p_j^*,\, j\in \Gamma)$, such that
\begin{equation}
\label{Wstar}
\sum_{m\in \Gamma}A_{nm}(x)p^*_m(x)=0,\quad \forall n \in \Gamma.
\end{equation}
The above stochastic model defines a one-dimensional PDMP. 

Let $X(t)$ and $\sigma(t)$ denote the stochastic continuous and discrete variables, respectively, at time $t$, $t>0$, given the initial conditions $X(0)=x_0,\sigma(0)=\sigma_0$. Introduce the probability density $p_n(x,t|x_0,n_0,0) $ with
\[\P\{X(t)\in (x,x+dx),\, \sigma(t) =n|x_0,\sigma_0)=p_n(x,t|x_0,\sigma_0,0)dx.\]
It follows that $p$ evolves according to the forward differential Chapman-Kolmogorov (CK) equation
\begin{equation}
\label{CKH}
\frac{\partial p_n}{\partial t}={\mathbb L}p_n ,
\end{equation}
with the operator ${\mathbb L}$ (dropping the explicit dependence on initial conditions) defined according to
\begin{equation}
\label{linH}
{\mathbb L} p_n(x,t)=-\frac{\partial F_n(x)p_n(x,t)}{\partial x}+\sum_{m\in \Gamma}A_{nm}(x)p_m(x,t).
\end{equation}
The first term on the right-hand side represents the probability flow associated with the piecewise deterministic dynamics for a given $n$, whereas the second term represents jumps in the discrete state $n$.

For a given realization $\sigma_t$ define 
\begin{equation}
\calS(x_0,t,t_0,k)=\e^{k\int_{t_0}^t f(X_{\sigma_t}(s))ds},
\end{equation}
so that the associated generating function can be written as
 \begin{equation}
 \label{Qng}
 \calQ_{\sigma}(x_0,t,k)=\E[\calS(x_0,t,0,k)\big | 1_{\sigma(0)=\sigma}].
 \end{equation}
We proceed by first deriving a Feynman-Kac formula for $\calS$ and fixed $\sigma_t$, which takes the form of a stochastic Liouville equation. We then obtain the corresponding Feynman-Kac equation for $\calQ_{n}$ by averaging with respect to different realizations $\sigma_t$. This takes the form of a differential CK equation.

The first step is to introduce a path-integral representation of the sample paths $X_{\sigma}(t)$. First, discretize time by dividing the given interval $[0,t]$ into $N$ equal subintervals of size $\Delta t$ such that $t=N\Delta t$ and set $x_j=X_{\sigma}(j\Delta t),\sigma_j=\sigma(j\Delta t)$ for $j=0,\ldots,N$. The  probability density for $x_1,\ldots,x_N$ given a particular realization of the stochastic discrete variables $\sigma_j,j=0,\ldots,N-1$, is
\begin{eqnarray*}
&&P_{\sigma}(x_1,\ldots,x_N) =\prod_{j=1}^{N-1} \delta \left (x_{j+1}-x_j-F_{\sigma_j}(x_j)\Delta t \right ).
\end{eqnarray*}
We define a corresponding discretized version of $\calS$ by
\begin{align}
\calS^{(N)}(x_0,t,0,k)&= \int_{\R^{N}}\exp\left (k\sum_{j=1}^Nf(x_j)\Delta t \right )\nonumber \\
&\times P_{\sigma}(x_1,\ldots,x_N)\left [\prod_{j=1}^Ndx_j\right ].
\label{Qds}
\end{align}
Taking the continuum limit $\Delta t\rightarrow 0,N\rightarrow \infty$ such that $N\Delta t= t$ yields the formal path-integral representation of $\calS$:
\begin{align}
\label{Qs}
&\calS(x_0,t,0,k)=\int_{\R}\left \langle \exp\left (k\int_{0}^{t} f(x(s))ds\right )\right \rangle_{x(t_0)=x_0}^{x(t)=x}dx,
\end{align}
where
\begin{align}
&\left \langle \exp\left (k\int_{0}^{t} f(x(s))ds\right )\right \rangle_{x(t_0)=x_0}^{x( t)=x}\\
&= \int_{x(t_0)=x_0}^{x(t)=x} \exp\left (k\int_{0}^tf(x(s))ds\right ){\mathcal P}_{\sigma}[x]{\mathcal D}[x] ,\nonumber
\end{align}
and
\begin{align*}
&\int_{x(0)=x_0}^{x(t)=x}{\mathcal P}_{\sigma}[x]{\mathcal D}[x]\\
&\quad =\lim_{\Delta t\rightarrow 0,N\rightarrow \infty}\int_{\Sigma^{N}}P_{\sigma}(x_0,x_1,\ldots,x_N)\prod_{j=1}^{N-1}dx_j.
\end{align*}

In order to derive a Feynman-Kac equation for $\calS$ we take the initial time to be $t-\tau$, set $\bar{\sigma}(\tau)=\sigma(t-\tau)$ and consider how $\calS(x_0,t,t-\tau,k)$ varies under the shift $\tau\rightarrow \tau +\Delta \tau$, with the final condition $\calS(x_0,t,t,k)=1$.
That is, 
\begin{align*}
&\calS(x_0,t,t-\tau-\Delta \tau,k)\\
&= \int_{\R}dx \left \langle \exp\left (k\int_{t-\tau-\Delta \tau}^{t} f(x(s))ds\right )\right \rangle_{x(t-\tau-\Delta \tau)=x_0}^{x(t)=x}\\
&\approx \int_{\R}dx \left \langle \exp\left (k\int_{t-\tau}^{t} f(x(s))ds\right )\right \rangle_{x(t-\tau)=x_0+\Delta x_0}^{x(t)=x} \e^{k f(x_0)\Delta \tau }\\
&=\e^{k f(x)\Delta \tau}\calS(x_0+\Delta x_0,t,t_0,k).
\end{align*}
We have split the time interval $[t-\tau-\Delta \tau,t]$ into two parts $[t-\tau,t]$ and $[t-\tau-\Delta \tau,t-\tau]$ and introduced the intermediate state $x(t-\tau)=x_0+\Delta x_0$ with $\Delta x_0$ determined by $\Delta x_0=F_{\bar{\sigma}}(x_0)\Delta \tau$. Expressing $\Delta x_0$ in terms of $\Delta \tau$ and Taylor expanding with respect to $\Delta \tau$
yields the following PDE in the limit $\Delta \tau\rightarrow 0$:
\begin{align}
\label{backQ}
\frac{\partial \calS}{\partial \tau}=F_{\bar{\sigma}}(x_0)\frac{\partial \calS}{\partial x_0}+kf(x_0)  \calS.
\end{align}

The crucial next step is to note that Eq. (\ref{backQ}) is a stochastic partial differential equation (SPDE), since $\bar{\sigma}(\tau)$ is a discrete random variable that varies with $\tau$ according to a Markov chain with adjoint matrix generator ${\mathbf M}^{\top}$. Since $\calS$ is a random field with respect to realizations of the discrete Markov process $\bar{\sigma}(\tau)$, there exists a probability density functional $\varrho$ that determines the statistics of $\calS(x_0,t,t-\tau,k)$ for fixed $k,t$. The expectation  $\E[\calS 1_{n(0)=n}]$ then corresponds to a first moment of this density functional. Rather than dealing with the probability density functional directly, we follow our previous work \cite{Bressloff17} by spatially discretizing the piecewise deterministic backward SPDE (\ref{backQ}) using a finite-difference scheme, take expectations and then recover the continuum limit.

Introduce the lattice spacing $\ell$ and set $x_j=j\ell, \ell \in \Z$.
Let $\calS_j(\tau,k)=\calS(j\ell,t,t-\tau,k)$, $f_j=f(j\ell)$, and $F_{j,n}=F(j\ell,n)$, $j\in \Z$. Eq. (\ref{backQ}) then reduces to the piecewise deterministic ODE (for fixed $k,t$)
\begin{equation}
\frac{d\calS_i}{d\tau}=F_{i,n}\sum_{j\in \Z}K_{ij} \calS_j+kf_i\calS_i,\quad \mbox{if } \bar{\sigma}(\tau)=n
\end{equation}
with
\begin{equation}
K_{ij}=\frac{1}{\ell}[\delta_{i,j-1}-\delta_{i,j}]
\end{equation}
Let ${\bf S}(\tau,k)=\{\calS_j(\tau,k),\, j \in \Z\}$ and introduce the probability density 
\begin{equation}
 \mbox{Prob}\{{\bf S}(\tau,k)\in ({\bf S},{\bf S}+d{\bf S}), \bar{\sigma}(\tau)=n\}=\varrho_n({\bf S},\tau)d{\bf S},
\end{equation}
where we have dropped the explicit dependence on initial conditions. The resulting CK equation for the discretized piecewise deterministic PDE is 
\begin{eqnarray}
\frac{\partial \varrho_n}{\partial \tau}&=&-\sum_{i\in \Z}\frac{\partial}{\partial \calS_i}\left [F_{i,n}\left (\sum_{j\in Z}K_{ij}  \calS_j\right )\varrho_n({\bf S},\tau)\right ] \nonumber \\
&&+\sum_{m\in \Gamma}A^{\top}_{nm}\varrho_m({\bf S},\tau).
\label{CK0}
\end{eqnarray}
Since the Liouville term in the CK equation is linear in ${\bf S}$, we can derive a closed set of equations for the first-order (and higher-order) moments of the density $\varrho_n$.

Let 
\begin{equation}
\label{ohoh}
\calQ_{j,n}(\tau,k)=\E[\calS_j(\tau,k)1_{\bar{\sigma}(\tau)=n}]=\int \varrho_n({\bf S},\tau)\calS_jd{\bf S},
\end{equation}
where
\[\int {\mathcal F}({\bf S})d{\bf S}=\left [\prod_j\int_0^{\infty} d\calS_j\right ]{\mathcal F}({\bf S})\]
for any ${\mathcal F}$.
Multiplying both sides of Eq. (\ref{CK0}) by $\calS_j$ and integrating with respect to ${\bf S}$ gives (after integrating by parts and assuming that $\varrho_n({\bf S},\tau)\rightarrow 0$ as ${\bf S}\rightarrow \infty$)
\begin{equation}
\frac{d {\mathcal Q}_{j,n}}{d \tau}=F_{j,n}\sum_{l\in \Z}K_{jl}{\mathcal Q}_{l,n}-sU_j{\mathcal Q}_{j,n}+ \sum_{m\in \Gamma}A^{\top}_{nm}{\mathcal Q}_{j,m}.
\end{equation}
If we now retake the continuum limit $\ell\rightarrow 0$ and set
\begin{equation}
\label{swv}
{\mathcal Q}_{\sigma}(x_0,t,k)=\E[\calS(x_0,t,t-\tau,k)\big |1_{\bar{\sigma}(\tau)=\sigma}]_{\tau=t}
\end{equation}
then we obtain the system of equations
 \begin{eqnarray} 
  \label{swCK2}
  \frac{\partial \calQ_{\sigma}}{\partial t} &=&\L_k^{\dagger} \calQ_{\sigma}\\
  &=&F_{\sigma}(x_0)\frac{\partial \calQ_{\sigma}}{\partial x_0}+kf(x_0)\calQ_{\sigma}+\sum_{m\in \Gamma} A^{\top}_{{\sigma}m}(x_0) \calQ_m.\nonumber 
  \end{eqnarray}
This is the Feynman-Kac formula for the moment generator (\ref{Qng}).
In the above derivation, we have assumed that integrating with respect to ${\bf S}$ and taking the continuum limit commute. (One can also avoid the issue that ${\bf S}$ is an infinite-dimensional vector by carrying out the discretization over the finite domain $[-L,L]$, and taking the limit $L\rightarrow \infty$ once the moment equations have been derived.) 
Finally, in order to obtain the Feynman-Kac equation (\ref{swCK}) for the two-state RPT, we take  
\[\Gamma=\{-1,1\},\quad F_{\sigma}=\sigma v, \quad {\bf A}=\left (\begin{array}{cc} - k& k\\k&-k\end{array}\right ).
\]

\end{document}